# 1D Convolutional Neural Networks and Applications – A Survey

Serkan Kiranyaz[1], Onur Avci[2], Osama Abdeljaber[3], Turker Ince[4], Moncef Gabbouj[5], Daniel J. Inman[6]

[1] Professor, Department of Electrical Engineering, Qatar University, Qatar.
Email: mkiranyaz@qu.edu.qa

[2] Former Assistant Professor, Department of Civil Engineering, Qatar University, Qatar.
Email: oavci@vt.edu

[3] Research Associate, Department of Civil Engineering, Qatar University, Qatar.
Email: o.abdeljaber@qu.edu.qa

[4] Professor, Electrical & Electronics Engineering Department, Izmir University of Economics, Turkey.
Email: turker.ince@izmirekonomi.edu.tr

[5] Professor, Department of Signal Processing, Tampere University of Technology, Finland;
Email: moncef.gabbouj@tut.fi

[6] Professor, Department of Aerospace Engineering, University of Michigan, Ann Arbor, MI, USA.
Email: daninman@umich.edu

**Abstract**
During the last decade, Convolutional Neural Networks (CNNs) have become the *de facto* standard for various Computer Vision and Machine Learning operations. CNNs are feed-forward Artificial Neural Networks (ANNs) with alternating convolutional and subsampling layers. Deep 2D CNNs with many hidden layers and millions of parameters have the ability to learn complex objects and patterns providing that they can be trained on a massive size visual database with ground-truth labels. With a proper training, this unique ability makes them the primary tool for various engineering applications for 2D signals such as images and video frames. Yet, this may not be a viable option in numerous applications over 1D signals especially when the training data is scarce or application-specific. To address this issue, 1D CNNs have recently been proposed and immediately achieved the state-of-the-art performance levels in several applications such as personalized biomedical data classification and early diagnosis, structural health monitoring, anomaly detection and identification in power electronics and motor-fault detection. Another major advantage is that a real-time and low-cost hardware implementation is feasible due to the simple and compact configuration of 1D CNNs that perform only 1D convolutions (scalar multiplications and additions). This paper presents a comprehensive review of the general architecture and principals of 1D CNNs along with their major engineering applications, especially focused on the recent progress in this field. Their state-of-the-art performance is highlighted concluding with their unique properties. The benchmark datasets and the principal 1D CNN software used in those applications are also publically shared in a dedicated website.

# 1 Introduction

Artificial neurons used in conventional ANNs are the first-order (linear) models of biological neurons. In the mammalian nervous system, the biological learning is mainly performed at the cellular level. As shown in Figure 1, each neuron is capable of processing the electrical signal based on the three individual operations [1,2]: 1) reception of the other neurons outputs through the synaptic connections in *Dendrites*, 2) the integration (or pooling) of the processed output signals in the *soma* at the nucleus of the cell, and, 3) the activation of the final signal at the first part of the *Axon* or the so-called *Axon hillock*: if the pooled potentials exceed a certain limit, it "activates" a series of pulses (action potentials). As shown in Figure 1(b), each terminal button is connected to other neurons across a small gap called a *synapse*. During the 1940s the first "artificial neuron" model was proposed by McCulloch-Pitts [3], which has thereafter been used in various feed-forward ANNs such as Multi-Layer Perceptrons (MLPs). As expressed in Eq. (1), in this popular model the artificial neuron performs a linear transformation through a weighted summation by the scalar weights. So, the basic operations performed in a biological neuron, that operate the individual synaptic connections with specific neurochemical operations and the integration in the cell's *soma* are modeled as the linear transformation (linear weighted sum) followed by a possibly nonlinear thresholding function, *f*(.), which is called activation function.

$$x_k^l = b_k^l + \sum_{i=1}^{N_{l-1}} w_{ik}^{l-1} y_i^{l-1} \quad and \quad y_k^l = f(x_k^l) \tag{1}$$

The concept of "Perceptron", was proposed by Frank Rosenblatt in his seminal work [4]. When used in all neurons of a MLP, this linear model is a basic model of the biological neurons leading to well-known variations in learning and generalization performances for various problems [4–8]. In the literature, there have been some attempts to change MLPs by modifying the neuron model and/or the conventional Back Propagation (BP) algorithm [9–11], or the MLP configuration [12–14] or even the way to update the network parameters (weights and biases) [15]. The most promising variant is called Generalized Operational Perceptrons [7,8,16,17], which is a heterogeneous network with non-linear operators and has thus exhibited significantly superior performance than MLPs; however, this is still the most common network model that has inspired the modern-age ANNs that are being used today.

Starting from the 1959, Hubel and Wiesel have established the foundations of the visual neuroscience through the study of the visual cortical system of cats. Their collaboration has lasted more than 25 years during which they have described the major responsive properties of the visual cortical neurons, the concept of receptive field, the functional properties of the visual cortex and the role of the visual experience in shaping the cortical architecture, in a series of articles published in *The Journal of Physiology* [18–22]. They are the pioneers who found the hierarchical processing mechanism of information in the visual cortical pathway, which eventually led to the Nobel Prize in Physiology or Medicine in 1981. With these advances in neurocognitive science, Fukushima and Miyake [23] in 1982 proposed the predecessor of Convolutional Neural Networks (CNNs), at the time called as "Neocognitron" which is a self-organized, hierarchical network and has the capability to recognize stimulus patterns based on the differences in their appearances (e.g., shapes). This was the first network, which has the unique ability of a biological mammalian visual system, that is, the assessment of similar objects to be assigned to the same object category independent from their position and certain morphological variations. However, in an attempt to maximize the learning performance, the crucial need of a supervised method to train (or adapt) the network for the learning task in hand became imminent. The ground-breaking invention of the Back-Propagation (BP) by Rumelhart and Hinton in 1986 [24] became a major cornerstone of the Machine Learning (ML) era. BP incrementally optimizes the network parameters, i.e., weights and biases, in an iterative manner using the gradient descent optimization technique.

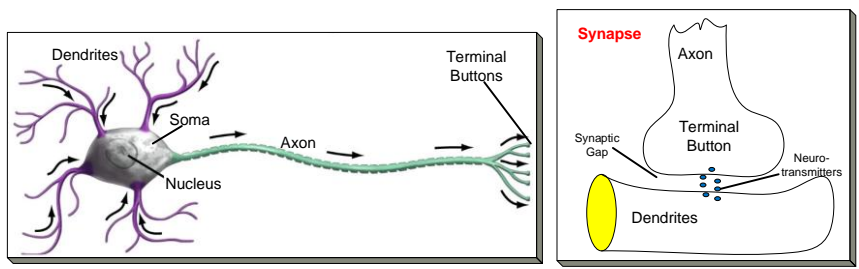

**Figure 1:** A biological neuron (left) with the direction of the signal flow and a synapse (right) [7].

These two accomplishments have started a new wave of approaches that eventually created the first naïve CNN models but it was the seminal work of Yann LeCun in 1990 who formulated the BP to train the first CNN [25], the so-called "LeNet". This CNN ancestor became mature in 1998 and its superior classification power was demonstrated in [26] over the benchmark MNIST handwritten number database [27]. This success has begun the era of CNNs and brought a new hope to otherwise "idle" world of ML during the 1980s and early 90s. CNNs have been used in many applications during the 90s and the first decade of the 21$^{st}$ century but soon they fell out of fashion especially with the emergence of new generation ML paradigms such as Support Vector Machines (SVMs) and Bayesian Networks (BNs). There are two main reasons for this. First, small or medium size databases were insufficient to train a deep CNN with a superior generalization capability. Then of course, training a deep CNN is computationally very demanding and feasible only with the modern graphical processors present today. This is why during these two decades the application of CNNs has been limited only to low-resolution (e.g. the thumbnail size) and gray-scale images in small-size datasets. On the contrary, both SVMs and BNs in comparison have fewer parameters that can be well optimized especially over such small to medium size datasets and independent from the image resolution. The annual ImageNet Large Scale Visual Recognition Challenge (ILSVRC) in the image classification competition in 2012 became the turning point for the application of deep CNNs in the area of large-scale image classification. For this competition, Krizhevsky *et al*. proposed the deep CNN model for the first time, the so-called AlexNet [28] which is the ancestor of the Deep Learning paradigm. AlexNet was an 8-layer CNN (5 convolutional-pooling layers and 3 fully-connected layers) that achieved 16.4% error rate in the ImageNet benchmark database [29] and this was about 10% lower than the second top method that uses a traditional ML approach, i.e., the Support Vector Machine (SVM) classifier network over the traditional visual features of Scale Invariant Feature Transform (SIFT) [30] and Local Binary Patterns (LBP) [31]. The ImageNet database contains more than one million images for training and divided into 1000 visual categories. The same study [28] also proposed some novel architectural features such as Rectified Linear Units (ReLU) instead of traditional activation functions such as Sigmoids (*sigm*) or Tangent Hyperbolics (*tanh*). The AlexNet team also proposed the Dropout method in [32] to improve the generalization capability of the network. However, the most important factor which made CNNs the mainstream method afterwards was the ability to train them over a massive size dataset by using parallelized computational paradigms over the emerging graphical processing units (GPUs).

With the successful introduction of AlexNet, the era of deep 2D CNNs has begun and immediately replaced the traditional classification and recognition methods within a short time. Deep CNNs eventually have become the primary tool used in any Deep Learning (DL) application including the contemporary ILSVRC image classification competitions. The following year, a new network, the so-called ZFnet [33] was proposed by Zeiler and Fergus that became the winning CNN model of the ILSVRC 2013. ZFnet further reduced the error rate down to 11.7% on the ImageNet database. The authors have shown how to visualize each convolution layer of the CNN which in turn has deepened our understanding why CNNs achieve such superior discrimination power among different visual object categories. The following year in ILSVRC 2014, a new breakthrough was achieved by the Google team with a deeper CNN, called as "GoogLeNet" with a codename "Inception", which almost halved the best error rate down to 6.7% in the ImageNet database. GoogLeNet has been designed by increasing the depth (with a 22 convolutional layers) and also the width of the network while keeping the computational budget constant. Besides using a deeper network with sparse connections, the key idea is that GoogLeNet obtained the top object recognition performance in ILSVRC 2014 with an ensemble of 6 CNNs. Since then, the popularity of the deep CNNs has peaked and eventually they became the *de facto* standard for various ML and computer vision applications over the years. Furthermore, they have been frequently used in processing sequential data including Natural Language Processing and Speech Recognition [34,35] and even 1D signals e.g., vibration [36,37].

Besides the top performance levels they can achieve, another crucial advantage they offer is that they can combine both feature extraction and classification tasks into a single body unlike traditional Artificial Neural Networks (ANNs). While conventional Machine Learning (ML) methods usually perform certain pre-processing steps and then use fixed and hand-crafted features which are not only sub-optimal but also require a high computational complexity, CNN-based methods learn to extract "optimized" features directly from the raw data for the problem at hand to maximize the classification accuracy. This is indeed the key characteristic for improving the classification performance significantly which made CNNs attractive to complicated engineering applications. However, the reign of traditional ML approaches was still unchallenged for 1D signals since deep CNNs were modeled and created specifically for 2D signals and their application was not straightforward for 1D signal signals especially when the data is scarce. The direct utilization of a deep CNN for a 1-D signal processing application naturally needs a proper 1D to 2D conversion. Recently, researchers have tried to use deep CNNs for fault diagnosis of bearings [36–44]. For this purpose, different conversion techniques have been utilized to represent the 1D vibration signals in 2D. A commonly used technique is to directly reshape the vibration signal into an $n \times m$ matrix called "the vibration image" [41]. Another technique was used in [39] where two vibration signals were measured using two accelerometers. Then, Discrete Fourier Transform (DFT) was applied, and the two transformed signals were represented in a matrix which can be fed into a conventional deep CNN. For electrocardiogram (ECG) beat classification and arrhythmia detection, the common approach is to first compute power- or log-spectrogram to convert each ECG beat to a 2D image [45,46]. However, there are certain drawbacks and limitations of using such deep CNNs. Primarily, it is known that they pose a high computational complexity which requires special hardware especially for training. Therefore, 2D CNNs are not suitable for real-time applications on mobile and low-power/low-memory devices. In addition, proper training of deep CNNs requires a massive size dataset for training to achieve a reasonable generalization capability. This may not be a viable option for many practical 1D signal applications where labeled data can be scarce.

To incorporate such drawbacks, in 2015 Kiranyaz et al. [47] proposed the first compact and adaptive 1D CNNs to operate directly on patient-specific ECG data. In a relatively short time, 1D CNNs have become popular with a state-of-the-art performance in various signal processing applications such as early arrhythmia detection in electrocardiogram (ECG) beats [47–49], structural health monitoring and structural damage detection [50–54], high power engine fault monitoring [55] and real-time monitoring of high-power circuitry [56]. Furthermore, two recent studies have utilized 1D CNNs for damage detection in bearings [57–60]. However, in the latter study conducted by Zhang et al. [60], both single and ensemble of deep 1D CNN(s) were created to detect, localize, and quantify bearing faults. A deep configuration of 1D CNN used in this study consisted of 6 large convolutional layers followed by two fully connected layers. Other deep 1D CNN approaches have been recently proposed by [61–64] for anomaly detection in ECG signals. These deep configurations share the common drawbacks of their 2D counterparts. For example, in [60], several "tricks" were utilized to improve the generalization performance of the deep 1D CNN such as data augmentation, batch normalization, dropout, majority voting, etc. Another approach to tackle this problem is to utilize the majority of the dataset for training which may not be feasible in some practical applications. In the study [60], more than 96% of the total data is used to train the deep network. With that, the assumption that such a large set of training data would be available may hinder the utilization of this method in practice. Therefore, in this article the focus is drawn particularly on compact 1D CNNs with few hidden layers/neurons, and their applications to some major engineering problems with the assumption that the labeled data is scarce, or application or device-specific solutions are required to maximize the detection and identification accuracy. The benchmark datasets and the principal 1D CNN software used in those applications are now publically shared in [65].

The rest of the paper is organized as follows. Section 2 provides a general background on adaptive and compact 1D CNNs with the formulation for Back-Propagation (BP) training. Section 3 presents popular engineering applications of the 1D CNNs. Section 4 presents a detailed computational complexity analysis of the 1D CNNs and the computational times of the competing methods on a sample application domain. Finally, Section 5 concludes the paper and suggests topics for future directions on 1D CNNs.

## 2 Overview of Convolutional Neural Networks

Deep Learning (DL) is the latest achievement of the Machine Learning era where it has arguably become the most crucial breakthrough and research hotspot. DL starts to govern our daily life, presenting such solutions that can only be imagined in science fiction movies just a decade earlier. Even before the introduction of the AlexNet, perhaps one can consider that this era has begun with the ground-breaking article published in the journal, *Science*, in 2006 by Hinton and Salakhutdinov [61], which explained the role of "the depth" of an ANN in machine learning. It basically points out the fact that ANNs with several hidden layers can have a powerful learning ability, which can further be improved with the increasing depth –or equivalently the number of hidden layers. Hence comes the term "Deep" learning, a particular ML branch, which can tackle complex patterns and objects in massive size datasets.

In this section, we shall begin with the fundamental tool of DL, the deep (and conventional) CNNs whilst explaining their basic features and blocks. We will briefly discuss the most popular deep CNNs ever proposed and then move on with the most recent CNN architecture, the 1D CNNs, which are focused solely on 1D signal and data repositories. The particular focus will be drawn on compact and adaptive 1D CNN models, which can promise certain advantages and superiorities over their deep 2D counterparts.

### 2.1 2D Convolutional Neural Networks

Although it has been almost 30 years after the first CNN was proposed, modern CNN architectures still share the common properties with the very first one such as convolutional and pooling layers. Also, besides few variations, the popular training method, the Back-Propagation technique is another commonality since 90s. This section will provide a brief overview of the conventional deep CNNs while introducing the most fundamental ideas and cornerstone architectures of the past.

To start with, the popularity and the wide range of application domains of deep CNNs can be attributed to the following advantages:
1. CNNs fuse the feature extraction and feature classification processes into a single learning body. They can learn to optimize the features during the training phase directly from the raw input.
2. Since CNN neurons are sparsely-connected with tied weights, CNNs can process large inputs with a great computational efficiency compared to the conventional fully-connected Multi-Layer Perceptrons (MLP) networks.
3. CNNs are immune to small transformations in the input data including translation, scaling, skewing, and distortion.
4. CNNs can adapt to different input sizes.

In a conventional MLP, each hidden neuron contains scalar weights, input and output. However, due to the 2D nature of images, each neuron in a CNN contains 2-D planes for weights, which is known as the kernel, and input and output which is known as the feature map. Figure 2 illustrates the basic blocks of a conventional CNN that classifies a 24×24-pixel grayscale image into two categories. This sample network consists of two convolution and two pooling layers. The output of the last pooling layer is processed by a single fully-connected layer and followed by the output layer that produces the classification output. The interconnections feeding the convolutional layers are assigned by weighting filters ($w$) having a kernel size of $(K_x, K_y)$. The convolution takes place within the image boundaries; therefore, the feature map dimension is reduced by the $(K_x - 1, K_y - 1)$ pixels from the width and height, respectively. The subsampling factors $(S_x, S_y)$ are set in advance in the pooling layers. In the sample illustration in the figure, the kernel sizes corresponding to the two convolution layers were set to $K_x = K_y = 4$, while the subsampling factors are set as $S_x = S_y = 3$ for the first pooling layer and $S_x = S_y = 4$ for the second one. Note that these values were deliberately selected so that the outputs of the last pooling layer (i.e. the input to the fully-connected layer) are scalars. (1x1). The output layer consists of two fully-connected neurons corresponding to the number of classes to which the image is categorized. The following steps describe a complete forward-propagation process in this sample CNN:

1. A 24×24-pixel grayscale image is fed to the input layer of the CNN.
2. Each neuron of the 1st convolution layer performs a linear convolution between the image and corresponding filter to generate the input feature map of the neuron.
3. The input feature map of each neuron is passed through the activation function to generate the output feature map of the neuron of the convolution neuron.
4. In the pooling layer, each neuron's feature map is created by decimating the output feature map of the previous neuron of the convolution layer. In this example, 7×7 feature maps are created in the 1st pooling layer.
5. Steps 3 and 4 are repeated and the outputs of the 2nd pooling layer become the inputs of the fully-connected layers, which are identical to the layers of a conventional MLP.
6. The scalar outputs are forward-propagated through the following fully-connected and output layers to produce the final output the represents the classification of input image.

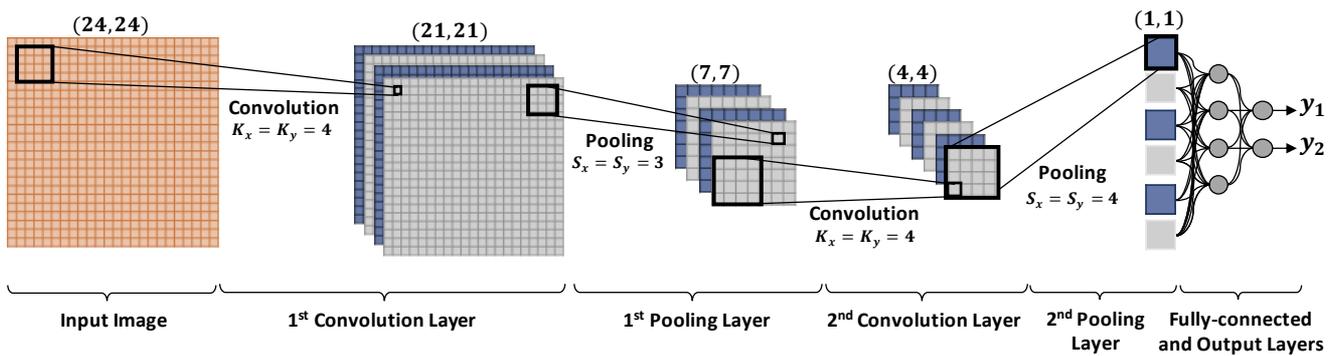

**Figure 2:** The illustration of a sample CNN with 2 convolution and one fully-connected layers.

CNNs are predominantly trained in a supervised manner by a stochastic gradient descent method, or the so-called backpropagation (BP) algorithm. During each iteration of the BP, the gradient magnitude (or sensitivity) of each network parameter such the weights of the convolution and fully-connected layers is computed. The parameter sensitivities are then used to iteratively update the CNN parameters until a certain stopping criterion is achieved. A detailed description of the BP in 2D CNNs can be found in [48].

The configurations of the two most popular CNNs, the ancestor CNN, "LeNet" [26] and the first deep CNN, "AlexNet" [28] are shown in Figure 3 and Figure 4, respectively. Despite the long time between them, the conceptual and architectural similarities are obvious. Perhaps the most striking difference is the "deep" configuration of the AlexNet, which encapsulates millions of network parameters. The fundamental two properties of the convolutional layers, "weight sharing" and "limited connectivity" exist even in the most-recent CNN architectures. In fact, these are the main features which separate CNNs from the conventional MLPs; otherwise, both networks are homogenous and share the common linear neuron model.

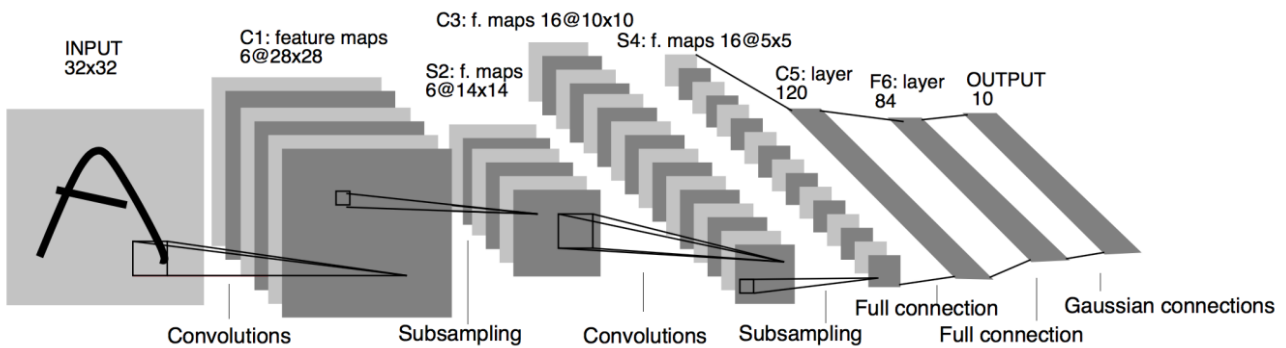

**Figure 3:** The configuration of the ancestor of CNNs, the "LeNet". The figure is taken from [26]. There are two interleaved convolutional and pooling layers following by three (two hidden and one output) fully-connected layers. The output layer is composed of 10 Radial Basis Function (RBF) neurons each of which computes the Euclidean distance between the network output and ground truth label for 10 classes.

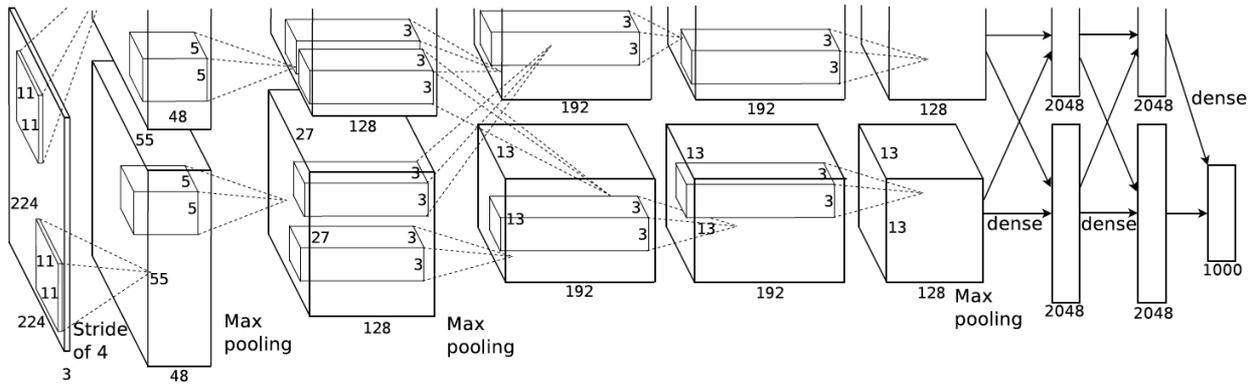

**Figure 4:** The configuration of the first deep CNN, the "AlexNet" [28]. There are 5 convolutional layers and 3 max-pooling layers following by three (two hidden and one output) fully-connected (dense) layers. The numbers of both convolution and fully-connected layers are significantly higher than the LeNet. The neurons at the output layer use softmax loss of the network predictions for 1000 classes.

## 2.2   1D Convolutional Neural Networks

The conventional deep CNNs presented in the previous section are designed to operate exclusively on 2D data such as images and videos. This is why they are often referred to as, "2D CNNs". As an alternative, a modified version of 2D CNNs called 1D Convolutional Neural Networks (1D CNNs) has recently been developed [47–56]. These studies have shown that for certain applications 1D CNNs are advantageous and thus preferable to their 2D counterparts in dealing with 1D signals due to the following reasons:

- Rather than matrix operations, FP and BP in 1D CNNs require simple array operations. This means that the computational complexity of 1D CNNs is significantly lower than 2D CNNs.
- Recent studies show that 1D CNNs with relatively shallow architectures (i.e. small number of hidden layers and neurons) are able to learn challenging tasks involving 1D signals. On the other hand, 2D CNNs usually require deeper architectures to handle such tasks. Obviously, networks with shallow architectures are much easier to train and implement.
- Usually, training deep 2D CNNs requires special hardware setup (e.g. Cloud computing or GPU farms). On the other hand, any CPU implementation over a standard computer is feasible and relatively fast for training compact 1D CNNs with few hidden layers (e.g. 2 or less) and neurons (e.g. < 50).
- Due to their low computational requirements, compact 1D CNNs are well-suited for real-time and low-cost applications especially on mobile or hand-held devices.

In the aforementioned recent studies, compact 1D CNNs have demonstrated a superior performance on those applications which have a limited labeled data and high signal variations acquired from different sources (i.e., patient ECG, civil, mechanical or aerospace structures, high-power circuitry, power engines or motors, etc.). As illustrated in Figure 5, two distinct layer types are proposed in 1D CNNs: 1) the so-called "CNN-layers" where both 1D convolutions and sub-sampling (pooling) occur, and 2) Fully-connected layers that are identical to the layers of a typical Multi-layer Perceptron (MLP) and therefore called as "MLP-layers". The configuration of a 1D-CNN is formed by the following hyper-parameters:

1) Number of hidden CNN and MLP layers/neurons (in the sample 1D CNN shown in Figure 5, there are 3 and 2 hidden CNN and MLP layers, respectively).
2) Filter (kernel) size in each CNN layer (in the sample 1D CNN shown in Figure 5, filter size is 41 in all hidden CNN layers).
3) Subsampling factor in each CNN layer (in the sample 1D CNN shown in Figure 5, subsampling factor is 4)
4) The choice of pooling and activation functions.

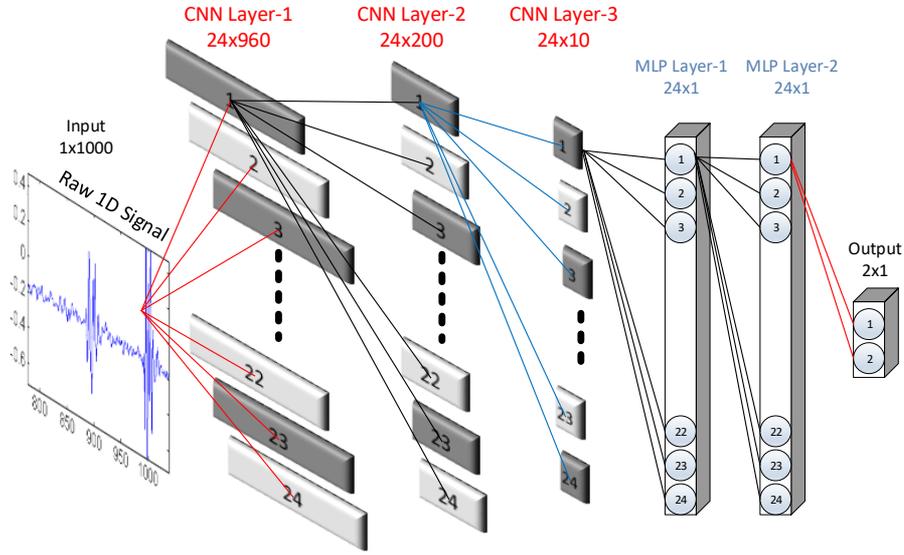

**Figure 5:** A sample 1D CNN configuration with 3 CNN and 2 MLP layers.

As in the conventional 2D CNNs, the input layer is a passive layer that receives the raw 1D signal and the output layer is a MLP layer with the number of neurons equal to the number of classes. Three consecutive CNN layers of a 1D CNN are presented in Figure 6. As shown in this figure, the 1D filter kernels have size 3 and the sub-sampling factor is 2 where the $k^{th}$ neuron in the hidden CNN layer, $l$, first performs a sequence of convolutions, the sum of which is passed through the activation function, $f$, followed by the sub-sampling operation. This is indeed the main difference between 1D and 2D CNNs, where 1D arrays replace 2D matrices for both kernels and feature maps. As a next step, the CNN layers process the raw 1D data and "learn to extract" such features which are used in the classification task performed by the MLP-layers. As a consequence, both feature extraction and classification operations are *fused* into one process that can be optimized to maximize the classification performance. This is the major advantage of 1D CNNs which can also result in a low computational complexity since the only operation with a significant cost is a sequence of 1D convolutions which are simply linear weighted sums of two 1D arrays. Such a linear operation during the Forward and Back-Propagation operations can effectively be executed in parallel.

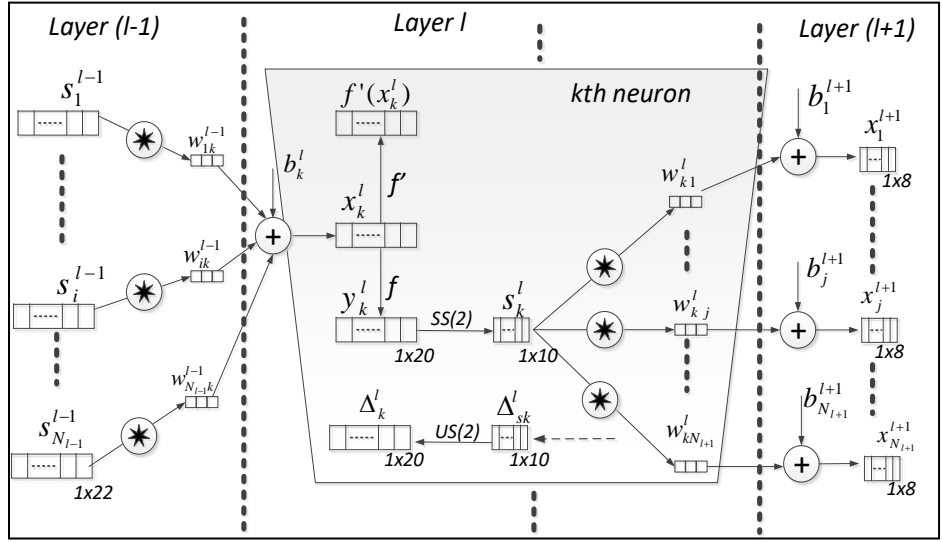

**Figure 6:** Three consecutive hidden CNN layers of a 1D CNN [56].

This is also an adaptive implementation since the CNN topology will allow the variations in the input layer dimension in such a way that the sub-sampling factor of the output CNN layer is tuned adaptively. The details related to Forward and Back-Propagation in CNN layers are presented in the next sub-section.

## 2.3 Forward- and Back-Propagation in CNN-layers

In each CNN-layer, 1D forward propagation (1D-FP) is expressed as follows:

$$x_k^l = b_k^l + \sum_{i=1}^{N_{l-1}} conv1D\left(w_{ik}^{l-1}, s_i^{l-1}\right) \quad (2)$$

where $x_k^l$ is defined as the input, $b_k^l$ is defined as the bias of the $k^{th}$ neuron at layer $l$, $s_i^{l-1}$ is the output of the $i^{th}$ neuron at layer $l-1$, $w_{ik}^{l-1}$ is the kernel from the $i^{th}$ neuron at layer $l-1$ to the $k^{th}$ neuron at layer $l$. $conv1D(.,.)$ is used to perform 'in-valid' 1D convolution without zero-padding. Therefore, the dimension of the input array, $x_k^l$, is less than the dimension of the output arrays, $s_i^{l-1}$. The intermediate output, $y_k^l$, can be expressed by passing the input $x_k^l$ through the activation function, $f(.)$, as,

$$y_k^l = f(x_k^l) \quad \text{and} \quad s_k^l = y_k^l \downarrow ss \tag{3}$$

where $s_k^l$ stands for the output of the $k^{th}$ neuron of the layer, $l$, and "$\downarrow ss$" represents the down-sampling operation with a scalar factor, $ss$.

The back-propagation (BP) algorithm can be summarized as follows. Back propagating the error starts from the output MLP-layer. Assume $l=1$ for the input layer and $l=L$ for the output layer. Let $N_L$ be the number of classes in the database; then, for an input vector $p$, and its target and output vectors, $\boldsymbol{t}^p$ and $[y_1^L, \cdots, y_{N_L}^L]'$, respectively. With that, in the output layer for the input $p$; the mean-squared error (MSE), $E_p$, can be expressed as follows:

$$E_p = \text{MSE}\left(\boldsymbol{t}^p, [y_1^L, \cdots, y_{N_L}^L]'\right) = \sum_{i=1}^{N_L}(y_i^L - t_i^p)^2 \tag{4}$$

To find the derivative of $E_p$ by each network parameter, the delta error, $\Delta_k^l = \frac{\partial E}{\partial x_k^l}$ should be computed. Specifically, for updating the bias of that neuron and all weights of the neurons in the preceding layer, one can use the chain-rule of derivatives as,

$$\frac{\partial E}{\partial w_{ik}^{l-1}} = \Delta_k^l y_i^{l-1} \quad \text{and} \quad \frac{\partial E}{\partial b_k^l} = \Delta_k^l \tag{5}$$

So, from the first MLP layer to the last CNN layer, the regular (scalar) BP is simply performed as,

$$\frac{\partial E}{\partial s_k^l} = \Delta s_k^l = \sum_{i=1}^{N_{l+1}} \frac{\partial E}{\partial x_i^{l+1}} \frac{\partial x_i^{l+1}}{\partial s_k^l} = \sum_{i=1}^{N_{l+1}} \Delta_i^{l+1} w_{ki}^l \tag{6}$$

Once the first BP is performed from the next layer, $l+1$, to the current layer, $l$, then one can carry on the BP to the input delta of the CNN layer $l$, $\Delta_k^l$. Let zero order up-sampled map be: $us_k^l = up(s_k^l)$, then the delta error can be expressed as follows:

$$\Delta_k^l = \frac{\partial E}{\partial y_k^l}\frac{\partial y_k^l}{\partial x_k^l} = \frac{\partial E}{\partial us_k^l}\frac{\partial us_k^l}{\partial y_k^l} f'(x_k^l) = up(\Delta s_k^l)\beta\, f'(x_k^l) \tag{7}$$

where $\beta = (ss)^{-1}$. Then, the BP of the delta error $\left(\Delta s_k^l \overset{\Sigma}{\leftarrow} \Delta_i^{l+1}\right)$ can be expressed as,

$$\Delta s_k^l = \sum_{i=1}^{N_{l+1}} conv1Dz\left(\Delta_i^{l+1}, rev(w_{ki}^l)\right) \tag{8}$$

where $rev(.)$ is used to *reverse* the array and $conv1Dz(.,.)$ is used to perform full 1D convolution with zero-padding. The weight and bias sensitivities can be expressed as follows:

$$\frac{\partial E}{\partial w_{ik}^l} = conv1D(s_k^l, \Delta_i^{l+1}) \quad \text{and} \quad \frac{\partial E}{\partial b_k^l} = \sum_n \Delta_k^l(n) \tag{9}$$

When the weight and bias sensitivities are computed, they can be used to update biases and weights with the learning factor, $\varepsilon$ as,

$$w_{ik}^{l-1}(t+1) = w_{ik}^{l-1}(t) - \varepsilon \frac{\partial E}{\partial w_{ik}^{l-1}} \quad \text{and} \quad b_k^l(t+1) = b_k^l(t) - \varepsilon \frac{\partial E}{\partial b_k^l} \tag{10}$$

The forward and back-propagation in hidden CNN layers are illustrated in Figure 7. The output sensitivity of the $k^{th}$ neuron at the CNN layer $l$, $\Delta \boldsymbol{s}_k^l$, is formed by back-propagating all the delta errors, $\Delta_i^{l+1}$, at the next layer, $\boldsymbol{l+1}$, by using Eq. (8), while the forward and back-propagation between the last hidden CNN layer and the first hidden MLP layer are summarized in Figure 8. Further details of the BP algorithm are presented in [48].

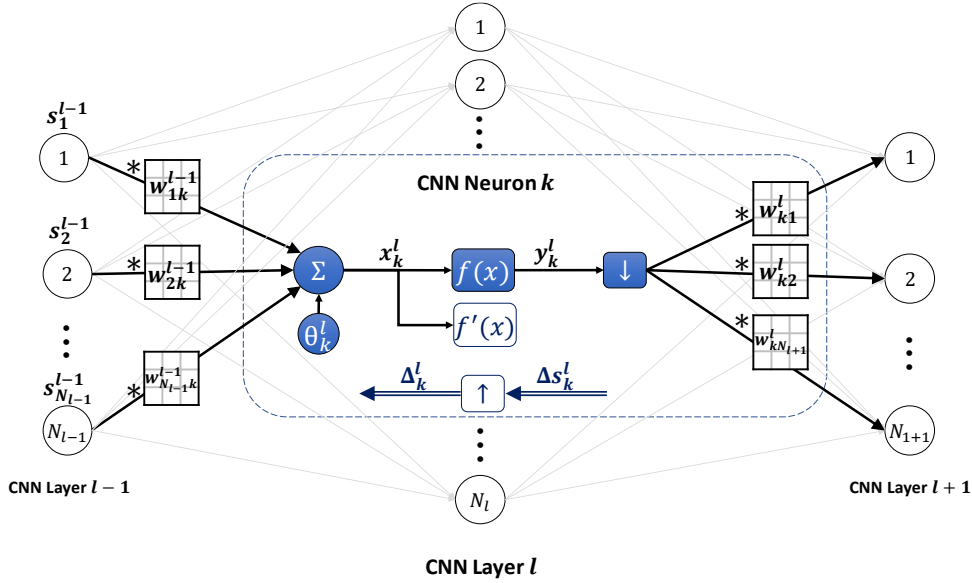

**Figure 7:** Forward and back-propagation in hidden CNN layers.

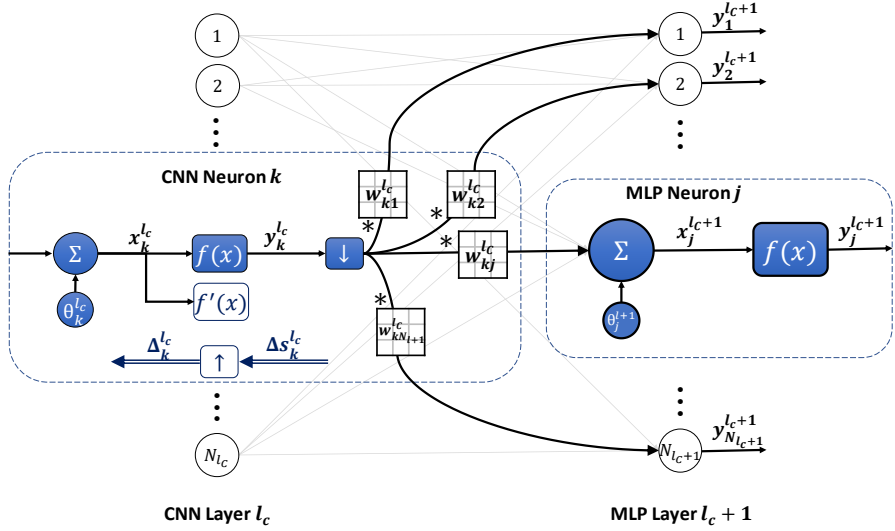

**Figure 8:** Forward and back-propagation between the last hidden CNN layer and the first hidden MLP layer.

Consequently, the iterative flow of the BP for the 1D raw signals in the training set can be stated as follows:

1) **Initialize** weights and biases (e.g., randomly, ~$U(-0.1, 0.1)$) of the network.
2) For each BP iteration **DO**:
    a. For each PCG beat in the dataset, DO:
        i. **FP**: Forward propagate from the input layer to the output layer to find outputs of each neuron at each layer, $s_i^l, \forall i \in [1, N_l]$, and $\forall l \in [1, L]$.
        ii. **BP**: Compute delta error at the output layer and back-propagate it to first hidden layer to compute the delta errors, $\Delta_k^l, \forall k \in [1, N_l]$, and $\forall l \in [1, L]$.
        iii. **PP**: Post-process to compute the weight and bias sensitivities using Eq. (9).
        iv. **Update**: Update the weights and biases by the (accumulation of) sensitivities scaled with the learning factor, $\varepsilon$ using Eq. (10).

## 3 Applications of 1D CNNs

As discussed earlier, the widespread use of CNNs is mainly motivated by their inherent capability to fuse feature extraction and classification into a single adaptive learning body. Due to the aforementioned reasons, there are several application domains where compact 1D CNNs have now been preferred over their 2D deep counterparts. Some of the major engineering applications of 1D CNNs will be presented in this section.

## 3.1 Real-time Electrocardiogram (ECG) Monitoring

While cardiovascular diseases are one of the major causes of deaths on the planet, irregular (arrhythmic) heartbeats have been reported to be an indication of a cardiovascular problem. Electrocardiogram (ECG) signals are extensively used by medical practitioners to monitor and evaluate the cardiac health. During the acquisition of an ECG, the electrical activity generated by the heart muscle is measured and displayed to detect cardiac abnormality; yet, the visual inspection and analysis of ECG signals by Cardiologists is laborious, subjective and wastes the crucial time for detecting a possible cardiac threat. The automated monitoring, detection and identification of the heart signals require a real-time, accurate and robust analysis that can be achieved by a dedicated ML approach. The first 1D CNN application was on ECG beat identification [14] where a "patient-specific" solution was proposed, i.e., for each arrhythmia patient a dedicated compact 1D CNN was trained by using the patient-specific training data as illustrated in Figure 9. The purpose is to identify each ECG beat into one of the five classes: N (beats originating in the sinus mode), S (supraventricular ectopic beats), V (ventricular ectopic beats), F (fusion beats), and Q (unclassifiable beats). In this study, ECG records from the benchmark MIT/BIH arrhythmia database [27] were used for both training and performance evaluation. There are a total of 48 records in this benchmark database and each record has two-channel ECG signal for 30-min duration selected from 24-hour recordings of 47 individuals. A total of 83648 beats from all 44 records were used as test patterns for performance evaluation. The proposed method has achieved the highest average accuracies (99% for Ventricular Ectopic Beats (VEB) and 97.6% for Supraventricular Ectopic Beats (SVEB)) on arrhythmia detection with the minimal computational complexity.

Several studies on arrhythmia detection and identification have been proposed ever since, e.g., [46,62–64]. However, all such studies focused on ECG beat classification for cardiac patients and strictly require a certain duration of training samples (e.g. 5 minutes) containing both normal and abnormal beats of the patient. In the absence of abnormal beats, which is the case of a healthy person, such methods cannot be applied for the early detection of abnormal beats for an otherwise healthy person with no past history of cardiac problems. This is basically a "Chicken and Egg" problem where one needs a certain number of abnormal samples to learn their characteristics in order to discriminate them from normal beats. A recent study [49] addressed this crucial problem and proposed a "personalized" solution for the early detection of cardiac arrhythmia at the moment they appear on an otherwise "healthy" person. This became the first attempt to propose a personalized early detection of ECG anomalies and cardiac health monitoring. In the absence of real abnormal beats, this becomes a far more challenging problem than the patient-specific ECG beat classification. The key accomplishment in this work is that the common causes of cardiac arrhythmias are modeled by a set of filters and then they are used to synthesize appropriate potential abnormal beats of a healthy person as illustrated in Figure 10. Upon learning the healthy person's (real) normal beats and potential (synthesized) abnormal beats, the proposed system with 1D CNNs can then be used to detect any abnormal beat which may occur during monitoring. Without using the real abnormal beats in training, the proposed method has achieved accuracy level, $Acc = 80.1\%$ and false-alarm rate, $FAR = 0.43\%$. The average probability of missing the first abnormal beat, therefore, is 0.199. In addition, the average probability of missing all three consecutive abnormal beats is around 0.0079. As a result, detecting one or more abnormal beat(s) among the first three occurrences is highly probable ($> 99.2\%$).

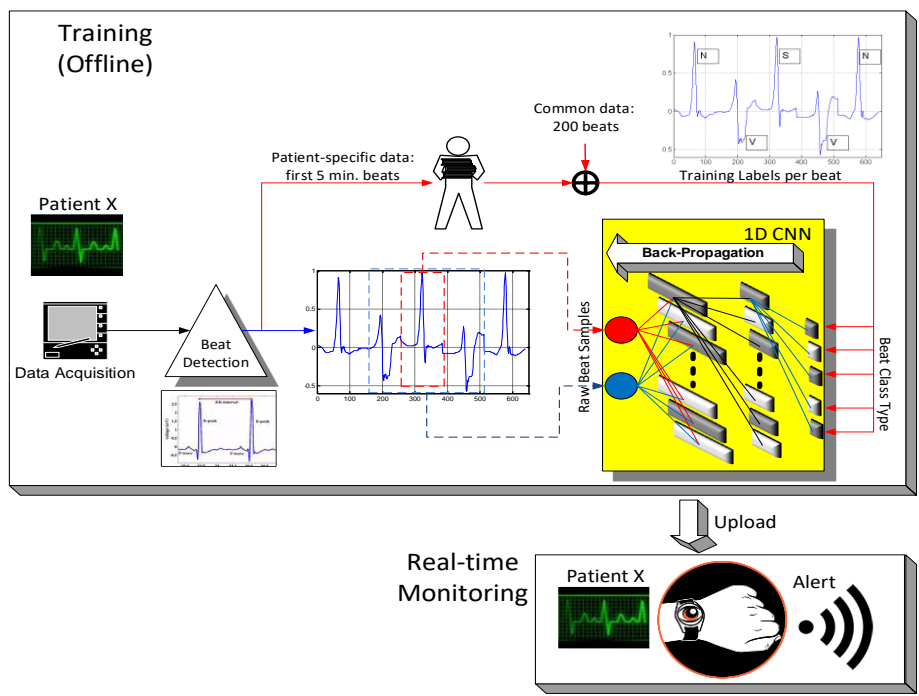

**Figure 9:** Overview of the arrhythmia detection and identification system proposed in [47,48].

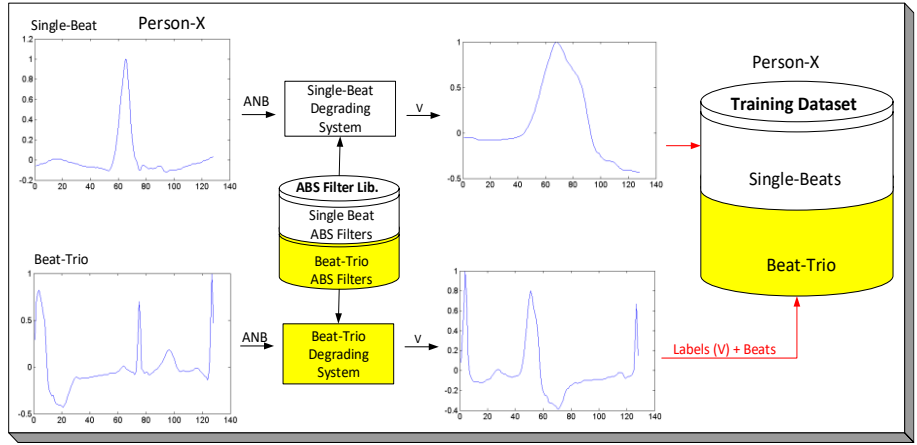

**Figure 10:** The creation of the training dataset as proposed in [49] for an arbitrary user (Person-X) using the real N-beats. The dataset is then used to train the dedicated 1D CNN for the user.

## 3.2 Vibration-Based Structural Damage Detection in Civil Infrastructure

Monitoring of structural damage is extremely important for sustaining and preserving the service life of civil engineering structures. Meticulous and early damage detection is one of the major concerns of structural health monitoring applications in civil engineering. While successful monitoring provides resolute and staunch information on the health, serviceability, integrity and safety of structures; maintaining high performance of a structure depends highly on monitoring the occurrence, formation and propagation of damage. Damage may accumulate on structures due to different environmental and human-induced factors. Numerous monitoring and detection approaches have been developed to provide practical means for early warning against structural damage. Tremendous efforts have been put into vibration-based damage detection methods which utilize the vibration response of the monitored structure to assess its health; and identify and locate structural damage. With emerging computing power, vibration-based techniques have become feasible, proved capable and widely used for structural damage detection.

Both parametric and nonparametric vibration-based damage detection methods that utilize conventional ML algorithms require extracting handcrafted features, which are then classified by an ML classifier. Numerous feature/classifier combinations have been examined in an attempt to find the best combination capable of optimally characterize the structural damage. Nevertheless, it is not guaranteed that a specific combination of feature and classifier will be suitable for all types of civil structures and all types of structural damage, at the same time. Moreover, the techniques usually used for extracting handcrafted features such as model identification, principle component analysis (PCA) and auto regressive (AR) modeling require significant computational time and effort.

Researchers have very recently started to apply novel DL algorithms to develop new vibration-based damage detection techniques that do not require manual feature extraction. Recent studies have shown that both 2D and 1D CNNs achieve superior performance levels in terms of their ability to detect and locate structural damage directly from the raw vibration signals without the need for data preprocessing or extraction of handcrafted features. Since 1D CNNs are easier to train and have lower computational complexity than their 2D counterparts, 1D CNNs are preferable when dealing with 1D vibration signals.

Conventional deep CNNs have been recently used to develop new techniques for vibration-based damage detection in civil structures. In a numerical study, Yu et al. [66] designed and trained a CNN to locate and quantify structural damage in a five-story structure. Since CNNs are only able to deal with 2D data (e.g. images), the 1D vibration signals acquired by 14 accelerometers were converted to a 2D representation simply by concatenating the 14 measured signals into a matrix. The data required for training the CNN was taken from a numerical model of the monitored structure under different damage scenarios. According to this dataset, a deep CNN having 3 CNN layers with a large number of neurons followed by 2 MLP layers was trained using a GPU implementation. The numerical results showed that the trained CNN was successful in detecting and locating simulated structural damage.

Khodabandehlou et al. [67] developed a similar CNN-based structural damage detection technique. A one-fourth–scale laboratory structure of a reinforced concrete bridge was used to experimentally demonstrate the proposed method. This structure was used to generate vibration data corresponding to four overall damage levels ranging from "no damage" to "extreme damage". For each damage level, the data measured by a number of accelerometers was concatenated into a single 2D matrix. The resulting dataset was used to train a deep CNN having 5 CNN and 4 MLP layers. It was demonstrated that the CNN was able to quantify the overall structural condition of the bridge directly from the measured vibration response.

In another study, Cha et al. [68] used a vision-based method with CNNs for detecting concrete cracks. It is reported that they achieved success in finding concrete cracks in realistic situations. In a similar study by Gulgec et al. [69] CNNs are used per a Python library Theano with the graphics processing unit (GPU) to classify damaged and undamaged samples modeled with Finite Element (FE) simulations only. It is reported that high classification accuracy is achieved for the FE data.

Abdeljaber et al. in [52] conducted 1D CNNs first time in vibration-based Structural Health Monitoring (SHM). A large-scale laboratory structure with plan dimensions of 5mx6m was constructed and instrumented with wired uniaxial accelerometers at Qatar University. The structure (QU grandstand simulator) is arguably the largest mock-up stadium structure built in a laboratory environment [70]. The steel structure has 30 joints at which vibrations response of the structure is recorded with accelerometers. The "damage" is introduced by simply loosening the bolts at a joint, which is an extremely slight change on the rotational stiffness as shown in Figure 6. The vibration response of the structure under 31 damage scenarios was measured and used to train an individual 1D CNN for each sensor location. Each 1D CNN was only responsible for processing the local data measured at the corresponding location. The performance of this 1D CNN-based damage detection method was tested under a large number of single and double damage cases. A complexity analysis was also conducted to estimate the computational time required for the 1D CNNs to process the measured signals. It was shown that the method was able to detect and locate structural damage in real-time. It was indeed the first time that compact 1D CNNs have proven to be able to accurately distinguish such uncorrelated and complex signals that can even defy a human expert inspector such as the two samples (damaged vs. undamaged) shown in Figure 11. Using an ordinary computer, when the performance of the proposed approach was tested, even with loosened bolts, all the damaged joints were detected without any misses or false alarms [52]. In addition, the detection speed was 45x faster than real-time speed. The 1D CNN application is optimized for multi-core CPU usage and can be obtained from [65]. This was an unprecedented achievement among all the CNN-based damage detection studies ever proposed.

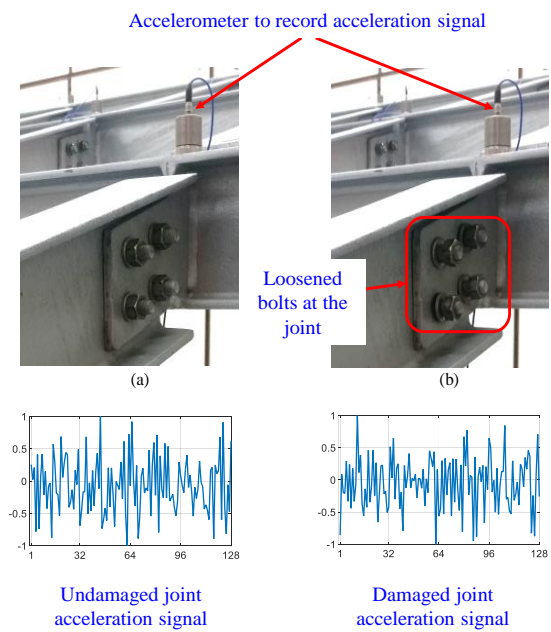

**Figure 11:** Damaged and undamaged conditions that are used in [52].

In an experimental study by Avci et al. [50], the 1D CNN-based method developed in [52] was integrated with a wireless sensor network (WSN). The method was modified to allow it to analyze the signals measured by the triaxial wireless sensors as shown in Figure 12. This was done so that the direction along which the damage-sensitive features are more pronounced can be determined. The modified damage detection technique was tested under a number of damage scenarios introduced to the laboratory structure. The results demonstrated the ability of the proposed technique to detect and localize damage directly from the ambient vibration response of the structure. All 1D CNNs trained in this study had a shallow structure (two CNN layers with only 4 neurons followed by two MLP layers with 5 neurons).

It was noticed that the process of generating the data required to train the 1D CNNs in [50,52,71] requires a large number of measurement sessions especially for a large civil structure. Therefore, Avci et al. in [53] and then Abdeljaber et al. in [54] developed a novel approach based on 1D CNNs, which require significantly less effort and labeled data for training. This approach was successfully tested over the data provided under the Experimental Phase II of the SHM Benchmark Problem [72].

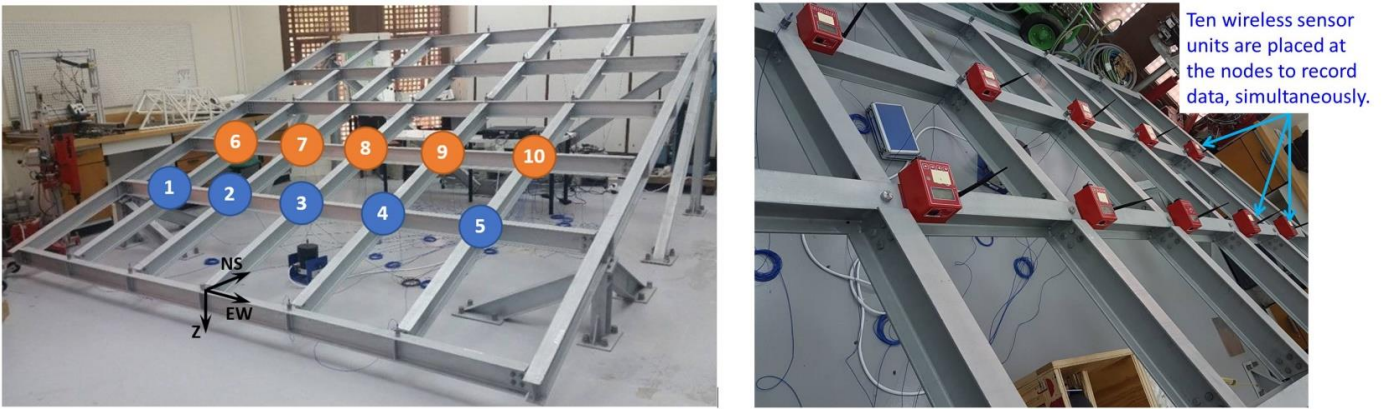

**Figure 12:** The test setup and wireless sensors used in [50,71].

### 3.3 Condition Monitoring in Rotating Mechanical/Aerospace Machine Parts

Engineers and scientists have been working on automatic solutions to accurately detect and identify any damage on various machine parts ranging from relatively small devices to large industrial systems. The crucial challenge in the latest research has been to determine whether the damage is at a critical level or it would propagate to reach a critical level for the structure. This is basically the quantification (grading) of the damage so as to make a final decision about the operational health of the machine part. Similar to civil engineering structures, SDD methods are typically utilized within the context of mechanical and aerospace engineering. It is important to note that the term Condition Monitoring (CM) in mechanical/aerospace engineering is analogous to SDD in civil structures; yet, it specifically addresses damage detection in rotating machinery. Reliability and operational efficiency of rotating machinery have always been crucial in modern machinery applications as the technology gets more advanced in relatively shorter time periods. Bearings play an important role in continuous functioning of the machines by being a direct interface between stationary internal supports and rotating components. Not only local bearings but also all types of rotating members affect the overall dynamic behavior, running accuracy, reliability and service life of the global machine structure.

As the bearings are continuously exposed to short- and long-term damage during operation; aging is inevitable for these elements. Not only the wearing and tear out, but also due to the lack in handling, repair and maintenance practices, more bearings continue to fail, reducing efficiency and reliability of the entire production line; carrying the potential to cause catastrophic failures of machine parts. Early detection of bearing faults by real-time condition assessment via embedded sensors has the potential to enable replacement of the bearing parts, instead of expensive replacements of the entire machine group. Various studies are available in the literature on diagnosis, prognosis; defect and fault detection; and condition monitoring of bearings and bearing parts in rotating machinery. Different techniques, such as acoustic emission, infra-red thermography, oil analysis have been used to determine the degradation of bearings. While these methods reveal the existence of defects inside bearings; they are not capable of localizing the defect (e.g., the rolling element, the cage, outher or inner race). Among the available approaches, vibration-based fault detection is standing out to be the most effective and reliable approach in revealing, locating and quantifying the damage on the bearing elements. Among the vibration-based techniques, ML based approaches have been predominantly emphasized and used more often for fault diagnosis of rotational machine parts. These methods are predominantly in need of damage-sensitive feature extraction from the recorded signals to be able to train a classifier for condition assessment of the bearing. Most of the available methods carry certain drawbacks and limitations. Due to the increasing complexity of mechanical elements and loading mechanisms, the degradation pattern can result in an enlargement of the damaged area and also in multiplication of number and location of defective structural parts. As a matter of fact, as the number of cracks in defective teeth gears [73] or number of spalls in defective bearings increase, various extracted features can lose their capability to determine the internal degradation as the number of simultaneous defects increases. In a broader sense, the main drawback of the early ML based methods is the fact that they are highly dependent on hand-crafted features. Such features might be sub-optimal, which means that they cannot accurately characterize the measured vibration signal. As such, when a classifier is trained based on sub-optimal features, it tends to result in unsatisfactory classification performance, and therefore unreliable fault diagnosis results [74].

From this perspective, it is crucial to detect the anomaly as soon as it appears so as to avoid large-scale damages or even worse, fatal outcomes such as electric discharges or potential explosions of machines. Numerous studies based on ML paradigms have been proposed in this domain with varying performance levels. This simply indicates how important the choice of the right features for the characterization of the monitored electric signals (e.g. current or voltage). It is a well-known fact that those fixed and handcrafted set of features cannot effectively characterize any possible electric signal and hence for those cases where their discrimination suffers, the detection performance will deteriorate regardless from the classifier used. This is why they are unable to form a generic solution that can be utilized for any electric waveform or data. Similar to other applications, 1D CNNs have the unique capability to optimize both feature extraction and classification in a single learning body and naturally, the three recent studies [55,58,59], have shown that a real-time monitoring and instantaneous anomaly detection can be accomplished with a state-of-the-art accuracy level. In [55] a novel method based on compact 1D CNNs was proposed to detect a potential motor anomaly due to the bearing faults. Bearing faults are mechanical defects that cause slight variations at certain frequencies in the motor-current waveform. Similar to the vibration signals, as shown in Figure 13 it is almost impossible to detect them visually by manual inspection even with a spectral analysis. 1D CNNs can accomplish this with close to 100% accuracy thanks to the layered sub-band decomposition performed in their hidden CNN-layers. Figure 14 shows the ROC curves of the 1D CNN method in [55] against the conventional methods such as [75–78].

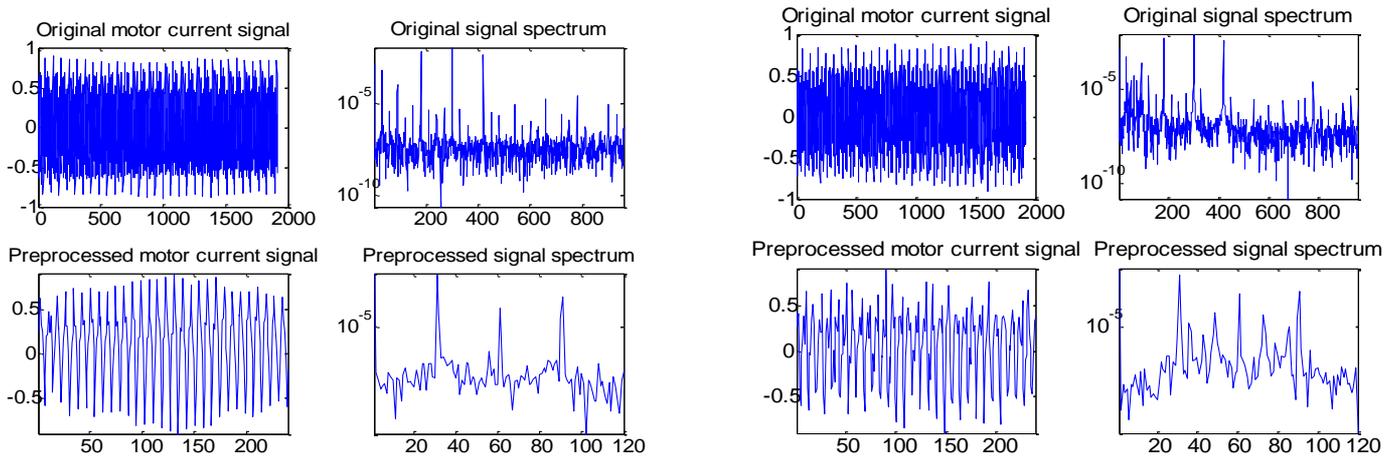

**Figure 13:** Sample healthy (left) and faulty (right) motor current signals and their spectrums before and after preprocessing [55].

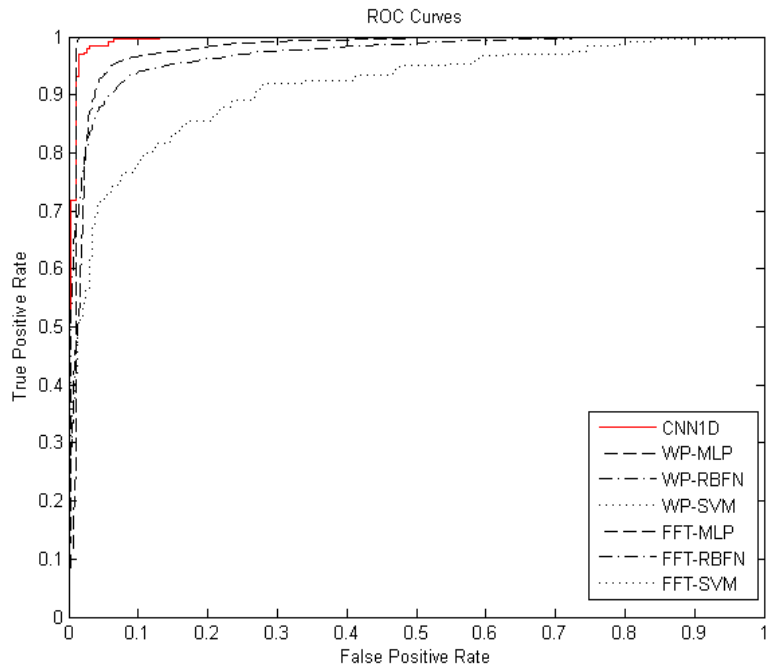

**Figure 14:** ROC plots of classifiers for comparison. The x- and y-axis represent the false positive rate and true positive rate, respectively [55].

## 3.4  Fault Detection in Modular Multilevel Converters (MMC)

High-power multilevel converters have been utilized extensively for efficient power conversion. The modular multilevel converters (MMC) are arguably the most efficient and feasible multilevel converter topology for medium power to high power applications. The MMC serves as a controllable voltage source with a high number of possible discrete voltage steps while the multilevel topology prevents potential major harmonic content generation. In comparison with other multilevel converter topologies, the predominant features of the MMCs are modularity and scalability to meet any voltage level requirements and performance efficiency in high-voltage applications. MMC is composed of many identical controlled voltage cells. Each cell can have one or more switches and a switch failure may occur in anyone of these cells. The steady-state normal and fault behavior of a cell voltage vary significantly based on the changes in the load current and the fault timing, which makes it difficult to detect and identify such faults in a fast manner. A conventional MMC as illustrated in Figure 15 provides a high power-voltage capability with a flexible control of the voltage level. Yet, safety and reliability has become the most important challenges for MMCs, which may encapsulate many power switching devices, each of which may be considered as a potential failure site. For instance, an open-circuit fault in a cell will distort the output voltage and current, which will cause an uncontrolled variation of the floating capacitor voltages and cause the disruption of operation and even a potential failure of the MMC.

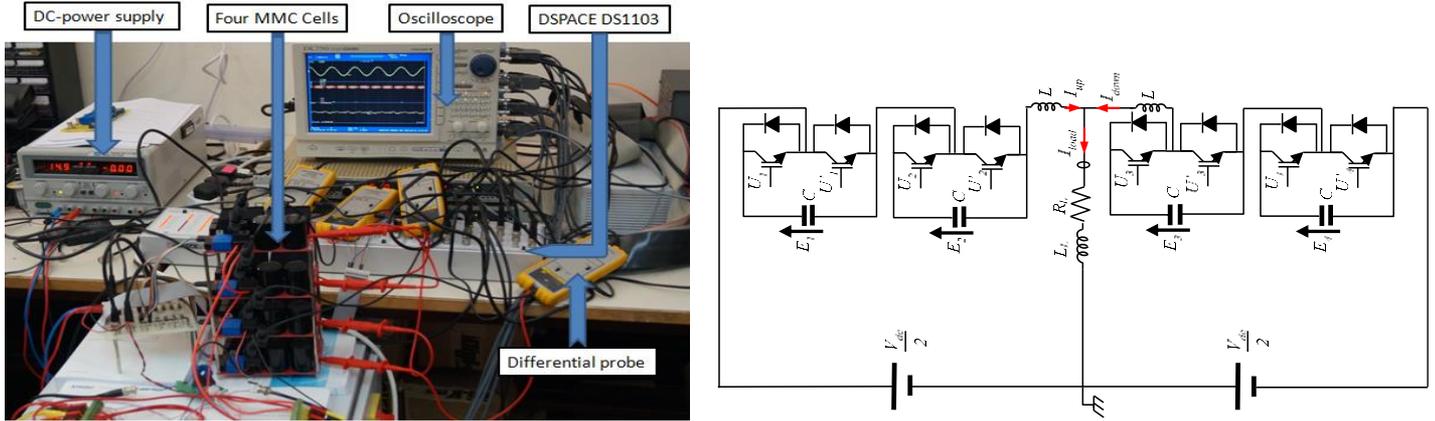

**Figure 15:** Implementation (left) and configuration (right) of the 4-cell MMC circuit [54].

Even though there are numerous studies for anomaly detection in MMC circuits, many of them contain limitations and drawbacks which may hinder the practical use of them. For example, some studies proposed to put sensor to each cell which may be neither feasible due to the high cost nor reliable since a sensor may fail too. Some other studies required manual feedback and human interaction. Most of them suffer from high computational complexity which hinders their utilization in real-time. The frontier study in [56] where a compact 1D CNN was used first time in the core of the system monitors the cell capacitor voltages and the differential current to detect an open-circuit anomaly almost instantaneously (Figure 16).

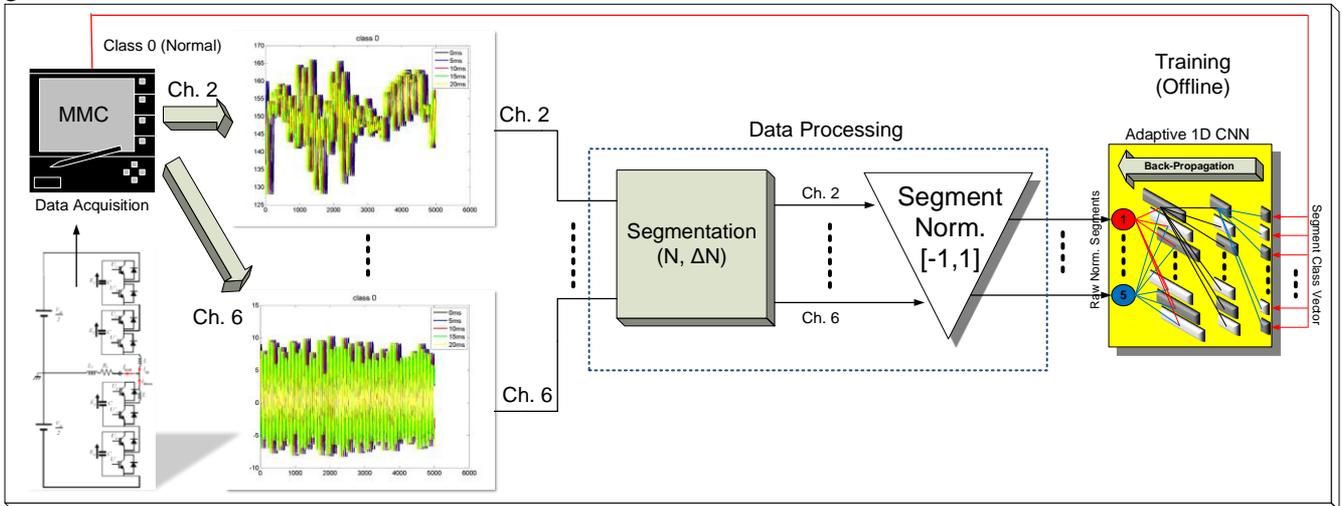

**Figure 16:** The main blocks of the proposed system in [56] and the offline training of the compact 1D CNN.

In summary, the following objectives have been accomplished by the proposed system:
1. Very high accuracy in fault detection and identification (e.g. practically 100%),
2. Excellent reliability and robustness against variations of MMC parameters and fault time,
3. Low computational complexity that allows real-time monitoring,
4. Low time delay for fault detection and identification (e.g., <0.1s), and
5. No false alarms.

Besides that, this method can easily scale up to very large scale MMCs with hundreds of cells and has the internal capability to detect the multiple faults. Such massive MMC circuits with hundreds or even thousands of SMs can be broken down to group of SMs with a practical size (e.g., 8 SMs) each of which can be monitored by a standalone and dedicated 1D CNN that can detect and localize any fault on that group -if and when occurs. Therefore, a group of identically trained 1D CNNs can monitor the entire MCM circuitry "in parallel" and if anyone detects a fault, the corresponding action (e.g. shutting of the source power) can immediately be taken. This is a straightforward expectation because the proposed 1D CNN detector has already shown the ability to distinguish the pattern of the real fault occurring on a particular switch from the other "distorted" patterns belonging to other switches functioning normally.

## 4 Computational Complexity Analysis of 1D-CNNs

In order to analyze the computational complexity for both FP and BP process, we shall first compute the total number of operations at each 1D CNN layer (ignoring the sub-sampling that has a negligible computational cost) and then cumulate them to find the overall computational complexity. During the FP, at a CNN layer, $l$, the number of connections to the previous layer is, $N^{l-1}N^l$, the number of connections to the previous layer is, $N^{l-1}N^l$, there is an individual linear convolution performed, which is a linear weighted sum. Let $sl^{l-1}$ and $wl^{l-1}$ be the vector sizes of the previous layer output, $s_k^{l-1}$, and kernel (weight), $w_{ki}^{l-1}$, respectively. Ignoring the boundary conditions, a linear convolution consists of $sl^{l-1}(wl^{l-1})^2$ multiplications and $sl^{l-1}$ additions from a single connection. Ignoring the bias addition, the total number of multiplications and additions in the layer $l$ will therefore be:

$$N(mul)^l = N^{l-1}N^l \; sl^{l-1}(wl^{l-1})^2,$$

$$N(add)^l = N^{l-1}N^l \; sl^{l-1}$$

(11)

So, during FP the total number of multiplications and additions, $T(mul)$, and $T(add)$, on a $L$ CNN layers will be,

$$T_{FP}(mul) = \sum_{l=1}^{L} N^{l-1}N^l \; sl^{l-1}(wl^{l-1})^2,$$

$$T_{FP}(add) = \sum_{l=1}^{L} N^{l-1}N^l \; sl^{l-1}$$

(12)

Obviously, $T(add)$ is insignificant compared to $T(mul)$.

During the BP, there are two convolutions performed as expressed in Eqs. (8) and (9). In Eq. (8) a linear convolution between the delta error in the next layer, $\Delta_i^{l+1}$, and the reversed kernel, $rev(w_{ik}^l)$, in the current layer, $l$. Let $xl^l$ be the size of both the input, $x_i^l$, and also its delta error, $\Delta_i^l$, vectors of the $i^{th}$ neuron. The number of connections between the two layers is, $N^{l+1}N^l$ and at each connection, the linear convolution in Eq. (8) consists of $xl^{l+1}(wl^l)^2$ multiplications and $xl^{l+1}$ additions. So, again ignoring the boundary conditions, during a BP iteration, the total number of multiplications and additions due to the first convolution will, therefore, be:

$$T_{BP}^1(mul) = \sum_{l=0}^{L-1} N^{l+1}N^l \; xl^{l+1}(wl^l)^2,$$

$$T_{BP}^1(add) = \sum_{l=0}^{L-1} N^{l+1}N^l \; xl^{l+1}$$

(13)

The second convolution in Eq. (9) is between the current layer output, $s_k^l$, and next layer delta error, $\Delta_i^{l+1}$ where $wl^l = xl^{l+1} - sl^l$. For each connection, the number of additions and multiplications will be, $wl^l$ and $wl^l(xl^{l+1})^2$, respectively. During a BP iteration, the total number of multiplications and additions due to the second convolution will, therefore, be:

$$T_{BP}^2(mul) = \sum_{l=0}^{L-1} N^{l+1}N^l \; wl^l \; (xl^{l+1})^2,$$

$$T_{BP}^2(add) = \sum_{l=0}^{L-1} N^{l+1}N^l \; wl^l$$

(14)

So at each BP iteration, the total number of multiplications and additions will be, $\left(T_{FP}(mul) + T_{BP}^1(mul) + T_{BP}^2(mul)\right)$ and $(T_{FP}(add) + T_{BP}^1(add) + T_{BP}^2(add))$, respectively. Obviously, the latter is insignificant compared to former especially when the kernel size is high. Moreover, both operation complexities are proportional to the total number of connections between two consecutive layers, which are the multiplication of the number of neurons at each layer. Finally, the computational complexity analysis of MLPs is well known (e.g., see [79]) and it is quite negligible in the current implementation since only a scalar (weight) multiplication and an addition are performed for each connection.

Without any exception, in all aforementioned 1D CNN applications, a minimal computational complexity is achieved against the competing (conventional) methods due to the aforementioned reasons. For example in the application of [52], using an ordinary computer, when the performance of the 1D CNN was tested, a fault detection speed of 45-times faster than real-time speed was achieved. As another example, Figure 17 presents the average classification times of the 1D CNN and the 6 competing methods for the application of motor fault detection where the competing methods are from [75–78].
.

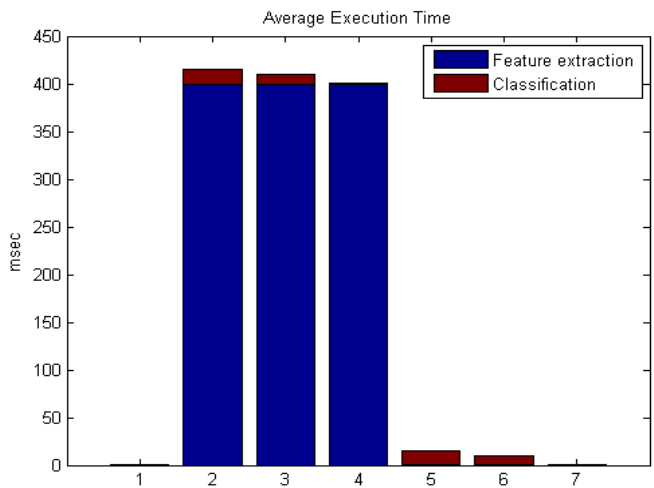

**Figure 17**: For the motor fault detection, the average execution times (in msec) of: (1) the 1D CNN method in [55], (2-7) 6 competing algorithms from [75–78].

## 5 Conclusions

Since the introduction of the first "artificial neuron" model by McCulloch-Pitts [3] in 1943, the era of ML has experienced many ups and downs in different stages of the history. Nevertheless, CNNs as its final product are attracting the utmost attention worldwide and influencing almost all aspects of the modern life. On one hand, its unique capability to learn directly from the raw signal or raw data voids the need of designing handcrafted pre-processing and feature extraction. On the other hand, deep CNNs alone can have an equal or even better learning ability than humans for the complex patterns or objects in massive size data repositories. Empowered by these, more and more Artificial Intelligence (AI) products are emerging every day, which will soon replace humans for many basic tasks such as driving, assisting, transportation, handling or load carrying, etc. It has already become apparent that AI will further assist or perhaps even replace humans on those complex tasks that require a high level of expertise and training, such as medical operations, health monitoring and diagnosis, taxonomy and even higher education.

1D CNNs are perhaps the latest success story of this era. Although they were introduced only a few years ago, recent studies have revealed that with a proper systematic approach, compact 1D CNNs can surpass all the traditional and conventional approaches. In this article, we draw the focus especially on those compact 1D CNNs and present a comprehensive survey on their engineering applications. Compact 1D CNNs can promise a sole advantage of being applicable to those applications where the labeled data for training is scarce and a low-cost, real-time implementation is desired. In such applications, it is obvious that a deep 2D CNN may not be feasible at all due to the scarcity of the training data and the high complexity that eventually violates the real-time constraint. On top of this, the conventional 2D CNNs can only process 2D signals; hence this enforces an extra 1D to 2D transformation following with a windowing (framing) operation, both of which cost additional time and resources. In many applications covered in this article, it has been shown that 1D CNNs are relatively easier to train and offer the minimal computational complexity while achieving state-of-the-art performance levels. They are especially suitable for mobile or hand-held devices with limited computation power and battery life. This is why they are attracting attention with an increasing pace; for instance, the 1D CNN publications, [48] and [52] have immediately become the most-popular and most-cited articles in their journals. The 1D CNN software used in these studies is now publically shared in [65].

The main limitation or the drawback of 1D CNNs is actually common for conventional CNNs and ANNs in general: They are homogenous (same neuron type in the entire network) and based solely on *linear-neuron* model from 1950s. A recent article [80] from an expert Neuroscientists pointed out this fact as follows: *"... But here's the thing: for all their similarities to the human brain, artificial deep neural nets are highly reductive models of the seemingly chaotic* *electro-chemical transmissions that populate every synapse of our own heads. With the big data era in neuroscience upon us, in which we can tease out the delicate wiring and diverse neuronal types (and non-neuron brain cells) that contribute to cognition, current deep learning models seem terribly over-simplistic.*" There are very recent attempts to address this deficiency of "modern-age" deep or compact ANNs. For instance, in studies [7,8], the first generalized neuron and network models, the so-called Generalized Operational Perceptrons (GOPs), have recently been proposed. GOPs can use *any* neuron model, linear or non-linear while having a heterogeneous network structure just like the human nervous system. Further in [16,17], GOPs were further improved to obtain other desired features such as neuron-level heterogeneity and "memory" capability. However, it was not until very recently that the first CNN-like network without those aforementioned limitations has been proposed in [81]. This new-generation network is called Operational Neural Networks (ONNs). ONNs can be heterogeneous and encapsulate neurons with any set of operators, linear or non-linear, to boost diversity and to learn highly complex and multi-modal functions or spaces with minimal network complexity and training data. These recent studies have shown that both GOPs and ONNs can indeed achieve significantly superior learning capabilities than the conventional MLPs and CNNs. Particularly, they exhibited an elegant learning performance over those highly complex and challenging problems which defy the conventional MLPs and CNNs. So, we can foresee that the emerging (1D) ONNs may soon replace 1D CNNs providing even more compact and efficient solutions especially for those 1D signal repositories with highly complex and diverse patterns.